\documentclass[cits,A4]{PoS}

\usepackage{amssymb}
\usepackage{amsmath}
\usepackage{amsfonts}
\usepackage{float}
\usepackage{cite}

\newcommand{\ep}{\varepsilon}
\newcommand{\N}{\nonumber}

\newcommand{\gsim}{\raisebox{-0.07cm   }
{$\, \stackrel{>}{{\scriptstyle\sim}}\, $}}

\title{{\footnotesize DESY 12-247, DO-TH 12/39, SFB/CPP 12-104, LPN12-141, \hfill 
{\tt arXiv:1212.5950[hep-ph]}}\\
New Results on the 3-Loop Heavy Flavor Wilson Coefficients in Deep-Inelastic 
       Scattering}

\ShortTitle{New Results on the 3-Loop Heavy Flavor Wilson Coefficients ...}

\author{Jakob Ablinger$^a$, Johannes Bl\"umlein$^b$, \speaker{Abilio De Freitas}$^b$, 
        Alexander Hassel- huhn$^{a,b}$, Sebastian Klein$^c$, 
        Carsten Schneider$^a$,~Fabian Wi{\ss}brock$^b$ \\
        \llap{$^a$} Research Institute for Symbolic Computation (RISC) Johannes Kepler 
        University, Altenbergerstra{\ss}e 69, A-4040 Linz, Austria \\
        \llap{$^b$} Deutsches Elektronen-Synchrotron, DESY, 
                    Platanenalle 6, D-15738 Zeuthen, Germany. \\
        \llap{$^c$} Institut f\"ur Theoretische Physik E, 
        RWTH Aachen University, D-52056 Aachen, Germany. \\}

\abstract{
\noindent
We report on recent results obtained for the 3-loop heavy flavor Wilson coefficients in 
deep-inelastic scattering (DIS) at general values of the Mellin variable $N$ at larger 
scales of $Q^2$. These concern contributions to the gluonic ladder-topologies, the 
transition matrix elements in the variable flavor scheme of $O(n_f T_F^2)$ and $O(T_F^2)$, 
and first results on higher 3-loop topologies. The knowledge of the heavy flavor Wilson 
coefficients at 3-loop order is of importance to extract the parton distribution 
functions and $\alpha_s(M_Z^2)$  in complete NNLO QCD analyses of the world precision 
data on the structure function $F_2(x,Q^2)$.}

\FullConference{36th International Conference on High Energy Physics\\
                 4-11 July 2012\\
                 Melbourne, Australia}

\begin{document}

The massive Wilson coefficients in deep-inelastic scattering are known to be expressible 
in the limit of high virtualities $Q^2 \gg m^2$ as convolutions between massive operator 
matrix elements (OMEs) and massless Wilson coefficients \cite{buza1}. Here $m$ denotes 
the heavy quark mass. The general structure of the Wilson coefficients to $O(\alpha_s^3)$
has been derived in \cite{bier1}. These massive Wilson coefficients are in turn  
convoluted with parton distribution functions to obtain the heavy flavor contributions to 
DIS structure functions at leading twist. They have been calculated for the twist-2 heavy 
flavor contributions to the unpolarized structure functions at leading \cite{witten1} and 
next-to-leading order \cite{laenen1} \footnote{For a precise implementation in Mellin 
space see \cite{Alekhin:2003ev}.} for general values of $Q^2$. Since the massless Wilson 
coefficients are known by now at 3-loop order \cite{vermaseren1}, it remains to compute 
the OMEs analytically at $O(\alpha_s^3)$, in order to obtain the massive Wilson coefficients
at NNLO. These coefficients will allow for a consistent NNLO analysis of the 
deep-inelastic world data at $Q^2 \gsim 20$GeV$^2$, cf.~\cite{Blumlein:2012bf}.

In these proceedings, we discuss recent progress obtained in this direction. Our aim is to 
calculate all contributing OMEs for general values of the Mellin variable $N$. An 
important previous step towards this goal was the computation of the moments of the massive
OMEs for $N=2 \ldots 10(14)$ contributing in the fixed and variable \footnote{In using 
variable flavor schemes  a correct scale matching is of importance 
\cite{Blumlein:1998sh}.} flavor schemes \cite{bier1}. The 3-loop heavy flavor 
corrections to $F_L(x,Q^2)$ in the asymptotic 
case were calculated in \cite{yo1}. First results for general values of $N$ for the 
color factor factors $T_F^2C_{A,F}$ were calculated in \cite{ablinger1} for two heavy 
quark lines of the same mass. The case of two different quark masses was 
considered in \cite{ablinger1,ablinger2} for fixed moments. Results for the color 
factors $n_fT_F^2C_{A,F}$ for general $N$ were obtained in\cite{ablinger3,blum1} and
the calculation of 3-loop ladder topologies was performed in \cite{ablinger4}. 
Two--loop results up to $O(\epsilon)$ were obtained in \cite{TL}.
Here the massive OMEs are computed for on-shell external massless partons. The case of 
a massive on-shell external fermion line was studied at two loops in \cite{yo2} in case of 
QED.

In the following we will describe the methods used to perform these computations. We 
generate the Feynman diagrams using {\tt QGRAF} \cite{qgraf}. After the numerators of 
these diagrams are contracted with appropriate projectors we end up with a large set of 
scalar integrals. Many of these integrals are calculated using a variety of
approaches, namely,
\begin{enumerate}
{\setlength \itemindent{4pt} \item  
Modern summation algorithms, implemented in the {\tt Mathematica} package {\tt Sigma} 
\cite{sigma}. 
}
{\setlength \itemindent{4pt} \item  The method of hyperlogarithms for 
convergent integrals, generalizing the method  developed \hspace*{1mm} in   
\cite{brown1} to one additional variable $x$.  
}
{\setlength \itemindent{4pt} \item  
Mellin-Barnes integral representations \cite{MB1}.}
{\setlength \itemindent{4pt} \item  
The use of integration by parts identities \cite{IBP1} to express all integrals in 
terms of a small set \hspace*{1mm} of masters integrals.}
\end{enumerate}

\begin{figure}[t]
\centering
\includegraphics[scale=1.1]{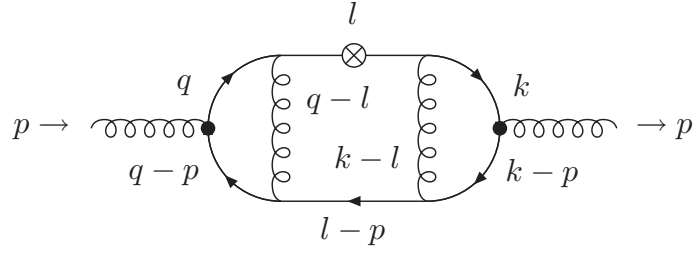} 
\caption{3-loop ladder diagram containing a central local operator insertion.}
\label{ladderA}
\end{figure}

We will focus here on the first two methods and show a few examples. The Feynman 
diagrams with operator insertions may be turned into nested sums \cite{REP}. These
infinite and finite sums  may be solved using {\tt Sigma} whenever they have a 
representation in terms of elements of difference- and product fields. This includes
divergent diagrams, since the different poles and powers in $\epsilon$ may be 
separated. Let us consider the scalar integrals associated with the ladder diagrams 
like the one shown in Fig.~\ref{ladderA}. In this diagram, the loop fermion 
is massive and the momentum of the external gluons is $p$, 
with $p^2=0$. We consider the case where all powers of propagators are equal to one, 
and in the numerator of the integral we only have the operator insertion 
$(\Delta \cdot l)^N$. The result after Feynman parameterization and calculation of 
the loop-momentum integrals  turns out to be \cite{ablinger4}
\begin{equation}
I_{1a} = \frac{i (\Delta . p)^N a_s^3 S_{\epsilon}^3}{(m^2)^{2-\frac{3}{2} \epsilon}} 
\hat{I}_{1a} \, ,
\end{equation}
where $S_{\epsilon}$ is the spherical factor 
$S_{\epsilon} = \exp \left[ \frac{\epsilon}{2}(\gamma_E -\ln(4 \pi)) \right]$, and
\begin{eqnarray}
\hat{I}_{1a} &=& -\exp \left( -\frac{3}{2} \epsilon \gamma_E \right) \Gamma(2-3\epsilon/2)
\prod_{i=1}^7 \int_0^1 dw_i 
\frac{\theta(1-w_1-w_2) w_1^{-\epsilon/2} w_2^{-\epsilon/2} (1-w_1-w_2)}{\left(1+w_1\frac{w_3}{1-w_3}+w_2\frac{w_4}{1-w_4}\right)^{2-3\epsilon/2}}
\nonumber \\ &&
\times w_3^{\epsilon/2} (1-w_3)^{-1+\epsilon/2} w_4^{\epsilon/2} (1-w_4)^{-1+\epsilon/2}
(1-w_5w_1-w_6w_2-(1-w_1-w_2)w_7)^N~.
\end{eqnarray}

Expanding the polynomial that appears raised to the $N$th power, one can see that the 
$w_1$- and $w_2$-integrals, can be written in terms of an Appell hypergeomteric 
function. After an appropriate analytic continuation, we end up with the following 
representation of the integral in terms of multiple sums,
\begin{eqnarray}
\hat{I}_{1a} &=&
\frac{\exp\left(-\frac{3}{2}\epsilon\gamma_E\right)\Gamma(2-3\epsilon/2)}{(N+1)(N+2)(N+3)}
\sum_{m,n=0}^{\infty} \left\{ \phantom{\sum_{t=1}^{N+2} \binom{N+3}{t}} \right. 
\nonumber \\ &&
\phantom{aaaaa} \sum_{t=1}^{N+2} \binom{N+3}{t}\frac{(t-\epsilon/2)_m 
(N+2+\epsilon/2)_{m+n}(N+3-t-\epsilon/2)_n}{(N+4-\epsilon)_{m+n}}
\nonumber \\ &&
\phantom{aaaaa \sum_{t=1}^{N+2}} \times 
\Gamma {t,t-\epsilon/2,m+1+\epsilon/2,n+1+\epsilon/2,N+3-t,N+3-t-\epsilon/2 \brack 
N+4-\epsilon,m+1,n+1,m+t+1+\epsilon/2,N+n-t+4+\epsilon/2}
\nonumber \\ &&
\phantom{aaaa} -\sum_{s=1}^{N+3} \sum_{r=1}^{s-1} \binom{s}{r} \binom{N+3}{s} (-1)^s
\frac{(r-\epsilon/2)_m (s-1+\epsilon/2)_{m+n} (s-r-\epsilon/2)_n}{(s+1-\epsilon)_{m+n}}
\nonumber \\ &&
\left. \phantom{aaaa -\sum_{s=1}^{N+3} \sum_{r=1}^{s-1}} \times
\Gamma {r,r-\epsilon/2,s-r,m+1+\epsilon/2,n+1+\epsilon/2,s-r-\epsilon/2 
\brack m+1,n+1,m+r+1+\epsilon/2,s-r+n+1+\epsilon/2,s+1-\epsilon} \right\}\, .
\end{eqnarray}
We can now expand in $\ep$, and the resulting multiple sums can then be performed using 
the package {\tt Sigma}. The result for this and other integrals can be written in 
terms of harmonic sums $S_{\vec{a}}$ \cite{HSUM} and their generalizations 
$S_{\vec{a}}(\vec{\xi};N)$ \cite{GENS1,GENS2} \footnote{Cyclotomic and generalized 
cyclotomic
harmonic sums and polylogarithms and their relations have been treated in 
\cite{CYCL}.}:
\begin{eqnarray}
S_{b,\vec{a}}(N) &=& \sum_{k=1}^N \frac{{\rm sign}(b)^k}{k^{|b|}} S_{\vec{a}}(k), \quad 
S_{\emptyset}(k)=1 \N\\
S_{b,\vec{a}}(\eta, \vec{\xi}; N) &=& \sum_{k=1}^N \frac{\eta^k}{k^b} 
S_{\vec{a}}(\vec{\xi};k), \quad S_{\emptyset}=1,  \quad \eta, \xi \in \mathbb{R} \, .
\end{eqnarray}
Omitting the explicit dependence of the harmonic sums on $N$, we obtain
\begin{eqnarray}
\hat{I}_{1a} &=& -\frac{4(N+1)S_1+4}{(N+1)^2(N+2)}\zeta_3 + \frac{2 S_{2,1,1}}{(N+2)(N+3)}
+ \frac{1}{(N+1)(N+2)(N+3)} \left\{ 
-2 (3N+5) S_{3,1} -\frac{S_1^4}{4} \right.
\nonumber \\ &&
+ \frac{4(N+1)S_1-4N}{N+1} S_{2,1}
+ 2 \left[ (2N+3)S_1+\frac{5N+6}{N+1} \right] S_3
+\frac{2 (3N+5) S_1^2}{(N+1)(N+2)}
+\frac{9+4N}{4}S_2^2 
\nonumber \\ &&
+ \left[ 2\frac{7N+11}{(N+1)(N+2)}+\frac{5N}{N+1}S_1-\frac{5}{2}S_1^2 \right] S_2
+\frac{N}{N+1} S_1^3 + \frac{4 (2N+3)}{(N+1)^2(N+2)}S_1
\nonumber \\ &&
-\frac{1}{2}(2N+3)S_4
\left. +8\frac{2N+3}{(N+1)^3(N+2)} \right\}~.
\end{eqnarray}
This result was checked using {\tt MATAD} \cite{matad} for the fixed moments 
$N=1 \ldots 10$. Other, more involved, integrals calculated in a similar way were given 
in Ref.~\cite{ablinger4}.

The second method we have used to compute the integrals is based on an algorithm 
originally proposed in \cite{brown1}. It is applicable when the integral turns out to 
be finite, even in case for local operator insertions for a fixed integer value of the
Mellin variable $N$. We have generalized this method to the case allowing for one 
non-vanishing fermion mass
and local operator insertions in order to find the general $N$-representations for  
convergent $3$-loop topologies. We work in the $\alpha$-representation and obtain 
integrals of the form
\begin{equation}
I_4(N)=\int \cdots \int d\alpha_1 ~d\alpha_2~ d\alpha_3~ d\alpha_4~ d\alpha_5~ 
d\alpha_6~ d\alpha_7~ d\alpha_8~ 
\frac{T}{U^2 V^2} \delta\left(1- \sum_i \alpha_i\right)\, .
\label{alphaInt}
\end{equation}
The corresponding graph polynomials of a graph $G$ are given by
\begin{itemize}
{\setlength \itemindent{8pt} \item $\phantom{.}$ $U=\sum_T \prod_{l \notin T} 
\alpha_l$, where $T$ denotes the spanning trees of $G$.}
{\setlength \itemindent{8pt} \item $\phantom{.}$ $V=\sum_{l \in massive} \alpha_l$.}
{\setlength \itemindent{8pt} \item $\phantom{.}$ Dodgson polynomials \cite{DODG} $T$ 
arise from the 
operator insertions. The form of these poly- \hspace*{4mm} nomials will 
depend on the 
specific operator insertion we are considering.}
\end{itemize}

The integrals given by (\ref{alphaInt}) are projective integrals, where one 
$\alpha$-parameter may be set to one eliminating the $\delta$--distribution.
The operators sit on on-shell diagrams which obey specific symmetries. These are generally 
not obeyed by the operator insertion. The Feynman parameter integrals are now performed 
in terms of hyperlogarithms \cite{brown1} $L(\overrightarrow{w},z): \mathbb{C} 
\backslash \Sigma \rightarrow \mathbb{C}$, where
\begin{itemize}
{\setlength \itemindent{8pt} \item $\phantom{.}$ 
$\Sigma=\{\sigma_0, \sigma_1,...,\sigma_N\}$ are distinct points in
$\mathbb{C}$ which may contain integration variables.}
{\setlength \itemindent{8pt} \item  $\phantom{.}$ $\overrightarrow{w}$ 
is a word over the alphabet $\mathfrak{A}=\{a_0,a_1,...,a_N\}$, where
    each letter $a_i$ corresponds to a \hspace*{4mm} point $\sigma_i$.}
\end{itemize}
$L\left({\overrightarrow{w}},z\right)$ is uniquely defined  by the following 
properties~:
\begin{eqnarray}
&& L\left(\{\},z\right)=1,~~\text{and}~~L\left({0^n},z,\right)=\frac{1}{n!} \log^n(z) 
~~\text{for}~~n \geq 1 \N\\
&&\frac{\partial}{\partial z} L\left(\left\{a_i \overrightarrow{w}\right\},z\right)=
\frac{1}{z-\sigma_i} L\left({\overrightarrow{w}},z\right)~~\text{for}~~z \in 
\mathbb{C} \backslash \Sigma \N\\
&&\text{If}~~\overrightarrow{w}~~\text{is not of the form}~~w=(0,0,\cdots,0), 
~~\text{then}
    \lim_{z\rightarrow 0}L\left({\overrightarrow{w}},z\right)=0.
\end{eqnarray}
For example, $L\left(a_i,z\right)=\log(z-\sigma_i)-\log(\sigma_i)$.

The hyperlogarithms satisfy shuffle relations, e.g.
\begin{equation}
 L\left(\{a_1,a_2\},z\right)
    L\left(\{a_3\},z\right)=L\left(\{a_3,a_1,a_2\},z\right)
 +L\left(\{a_1,a_3,a_2\},z\right)+L\left(\{a_1,a_2,a_3\},z\right)~.
\end{equation}
The points to which the indices $a_i$ correspond may contain further 
integration variables.
Using these properties after partial fractioning and integration
    by parts, one can express any primitive  for expressions consisting of rational and
    hyperlogarithmic functions in terms of different hyperlogarithmic
    functions. These primitives have to be evaluated at the respective integration limits.
Due to the operator-insertions leading  to power-type functions, the integrals do not fit directly
    into the framework of the algorithm for general values of $N$.
In order to obtain the corresponding extension
a generating function is constructed by the mapping,
\begin{equation}
p\left(\alpha_1,\cdots,\alpha_n\right)^N 
\rightarrow \sum_{k=0}^\infty x^k p\left(\alpha_1,\cdots,\alpha_n\right)^k
=  \frac {1} 
{1-x~p\left(\alpha_1,\cdots,\alpha_n\right)}~.
\label{eq:xmap}
\end{equation}
\begin{figure}[H]
\centering
\includegraphics[scale=1.2]{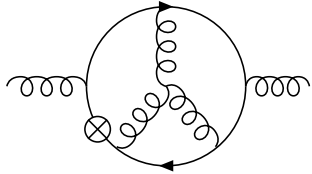}   
\caption{A 3-loop Benz diagram.}
\label{benzA}
\end{figure}

\noindent
Performing the Feynman-parameter integrations then leads to an expression which contains 
hyperlogarithms $L_w$  in the variable $x$.
Using this method, the scalar integral with all powers of propagators equal to one associated with the diagram
shown in Fig.~\ref{benzA}, corresponding to a Benz-type topology, yields
\begin{eqnarray}
I(x)&=& \frac{1} {(1+N) (2+N) x}
\Biggl\{
\zeta_3\Bigl[
{2}L\left(\{-1\},x\right)
-2 (-1+2 x) L\left(\{1\},x\right)
-4 L\left(\{1,1\},x\right)
\Bigr]
\nonumber \\ 
&&
-{3 } L\left(\{-1,0,0,1\},x\right)
+{2 }L\left(\{-1,0,1,1\},x\right)
-{2 x L\left(\{0,0,1,1\},x\right)}
+{3 x L\left(\{0,1,0,1\},x\right)}
\N\\
&&
-x {L\left(\{0,1,1,1\},x\right)}
+{(-3+2 x) L\left(\{1,0,0,1\},x\right)}
+{2 x L\left(\{1,0,1,1\},x\right)}
-{L\left(\{1,0,1,1,1\},x\right)}
\N
\end{eqnarray}
\begin{eqnarray}
&&
-{(5 x-1) L\left(\{1,1,0,1\},x\right)}
+x {L\left(\{1,1,1,1\},x\right)}
-{2 L\left(\{1,0,0,1,1\},x\right)}
+{3 L\left(\{1,0,1,0,1\},x\right)}
\nonumber \\ &&
+{2 L\left(\{1,1,0,0,1\},x\right)}
+{2 L\left(\{1,1,0,1,1\},x\right)}
-{5 L\left(\{1,1,1,0,1\},x\right)}
+{L\left(\{1,1,1,1,1\},x\right)}
\Biggr\}~.
\nonumber \\ &&
\end{eqnarray}
Finally, the $N$th coefficient of this expression in $x$ has to
be extracted analytically in order to undo the mapping (\ref{eq:xmap}).
This is achieved using the {\tt {GetMoments}} option of 
the package {\tt HarmonicSums} \cite{GENS2}. One may also use guessing-methods
to obtain the corresponding difference equation based on a huge number of moments,
cf.~\cite{Blumlein:2009tj}. For a more complicated graph with non-trivial argument 
structure in $x$ 
we were able to produce $\sim 1500$ moments \cite{ABSW13}. One obtains from Eq.~(10)
\begin{eqnarray}
I(N)&=&
\frac {1} {(N+1) (N+2) (N+3)}
\Biggl\{
\frac{648+1512 N+1458 N^2+744 N^3+212 N^4+32 N^5+2 N^6}{(1+N)^3
  (2+N)^3 (3+N)^3}
\nonumber \\ &&
-\frac{2 \left(-1+(-1)^N+N+(-1)^N N\right)}{(1+N)} \zeta_3
-(-1)^N S_{-3}
-\frac{N}{6 (1+N)} S_1^3
+\frac{1}{24} S_1^4
-\frac{1}{4} S_4
\nonumber \\ &&
-\frac{\left(7+22 N+10 N^2\right)}{2 (1+N)^2 (2+N)} S_2
-\frac{19}{8} S_2^2
-\frac{1+4 N+2 N^2}{2 (1+N)^2(2+N)} S_1^2
+\frac{9}{4} S_2
-\frac{(-9+4 N)}{3 (1+N)} S_3
\nonumber \\ &&
-{2 (-1)^N } S_{-2,1}
+\frac{(-1+6 N)}{(1+N)} S_{2,1}
+\frac{54+207 N+246 N^2+130 N^3+32 N^4+3 N^5}{(1+N)^3
    (2+N)^2 (3+N)^2} S_1
\nonumber \\ &&
+{4} \zeta_3 S_1
-\frac{(-2+7 N)}{2 (1+N)} S_2 S_1
+\frac{13}{3} S_3 S_1
-{7} S_{2,1} S_1
-{7} S_{3,1}
+{10} S_{2,1,1}
\Biggr\}~.
\end{eqnarray}

\begin{figure}[H]
\centering
\includegraphics[scale=1.2]{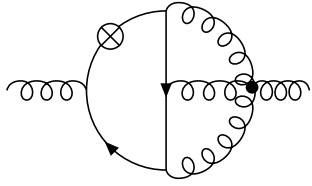} 
\caption{A second example of a 3-loop Benz topology.}
\label{benzB}
\end{figure}

Another example, where this technique has been applied, is shown in Fig.~\ref{benzB}.
In this case the result is
\begin{eqnarray}
I(N)&=&
\frac {1} {(N+1) (N+2)}
\Biggl\{
\frac{2 \left(1-13 (-1)^N+(-1)^N 2^{3+N}+N-7 (-1)^N N+3 (-1)^N
    2^{1+N}N\right) }{(1+N) (2+N)} \zeta_3
\nonumber \\&&
+\frac{1}{(2+N)} S_3
+\frac{(-1)^N}{2 (2+N)} S_1^3
-\frac{(-1)^N (3+2 N)}{2 (1+N)^2 (2+N)} S_2
+\frac{5 (-1)^N}{2} S_2^2
+\frac{2 (-1)^N (3+N) }{(1+N) (2+N)} S_{2,1}
\nonumber \\&&
+\frac{(-1)^N (3+2 N)}{2 (1+N)^2(2+N)} S_1^2
-\frac{(-1)^N}{2} S_2 S_1^2
+\frac{3 (-1)^N (4+3 N)}{(1+N) (2+N)} S_3
+3 (-1)^N S_4
+\frac{2}{(2+N)} S_{-2,1}
\N\\
&&
-{12 (-1)^N } S_1 \zeta_3
+\frac{(-1)^N (5+7 N) }{2 (1+N) (2+N)} S_1 S_2
+{3 (-1)^N } S_1 S_3
+{4 (-1)^N } S_{2,1} S_1
-{4 (-1)^N} S_{3,1}
\N
\end{eqnarray}
\begin{eqnarray}
&&
-\frac{4 \left((-1)^N 2^{2+N}-3 (-2)^N N+3 (-1)^N 2^{1+N} N\right)}{(1+N)
  (2+N)} S_{1,2}\left(\frac{1}{2},1\right)
-{5 (-1)^N } S_{2,1,1}
\nonumber \\&&
+{2 (-1)^N } \zeta_3 S_1 \left(2\right)
+\frac{2 \left(-(-1)^N 2^{2+N}-13 (-2)^N N+5 (-1)^N 2^{1+N}
    N\right)}{(1+N) (2+N)} S_{1,1,1}\left(\frac{1}{2},1,1\right)
\nonumber \\&&
-{2 (-1)^N } S_{1,1,2}\left(2,\frac{1}{2},1\right)
-{(-1)^N} S_{1,1,1,1}\left(2,\frac{1}{2},1,1\right)
\Biggr\} \, .
\end{eqnarray}

Notice the presence of generalized harmonic sums and highly divergent factors in the 
limit $N \rightarrow \infty$, 
such as $2^N$. It can be shown, however, that the complete expression is
convergent in this limit and possesses a well-defined asymptotic expansion for $N 
\rightarrow \infty$. In general neither the representation in individual nested sums or 
by iterated integrals shows this property, but a corresponding combination of terms 
does. 

\begin{figure}[H]
\centering
\includegraphics[scale=.22]{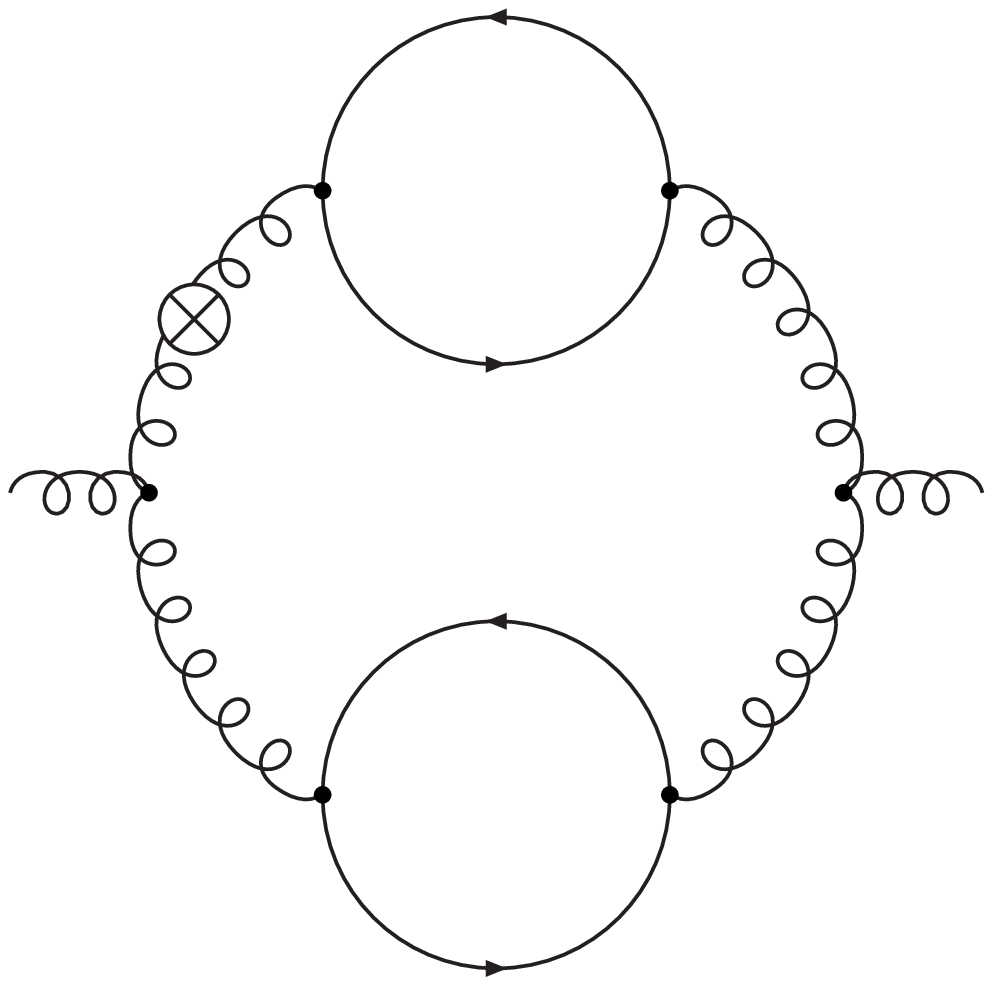} 
\caption{A 3-loop graph containing two massive fermion lines $m_1=m_2$ and an operator 
insertion.}
\label{Massive2.fig}
\end{figure}

We calculated the contributions of $O(n_f T_F^2 C_{A,F})$ to all massive OMEs
completely \cite{ablinger3,blum1}. Furthermore, first systematic results were
obtained for the case of graphs containing two massive fermion lines with $m_1 = m_2$.
A typical graph is shown in Fig.~\ref{Massive2.fig} in the gluonic case.

\noindent
One obtains
\begin{align}
 I =& 
  \frac{1}{105 \ep^2}
 +\frac{1 }{\ep}
  \Biggl[
    \frac{74 N^3-455 N^2+381 N-210}
    {44100 (N-1) N (N+1)}
   -\frac{1}{210} S_{1}(N)
  \Biggr]
\N\\&
 +\frac{8903 N^3+39537 N^2-114440 N+36576}
  {2822400 (N+1) (2 N-3) (2 N-1)} S_{1}(N)
\N\\&
 +\frac{P_1}
  {148176000 (N-1)^2 N^2 (N+1)^2 (2 N-3) (2 N-1)}
 +\frac{1}{840} 
  \Bigl( 
   S_{1}(N)^2 + S_{2}(N) + 3 \zeta_2
  \Bigr)
\N\\&
 +\frac{2^{-2 N-9} (N-1) N (5 N-6)}
  {3 (2 N-3) (2 N-1)}
  \binom{2N}{N}
  \Biggl(
   - 7 \zeta_3
   - \sum _{j=1}^N 
     \frac{4^{j} }
     {\binom{2j}{j} j^3}
   + \sum_{j=1}^N 
     \frac{4^{j} S_{1}(j)}
     {\binom{2j}{j} j^2}
   \Biggr)~,
\end{align}
\begin{align}
 P_1 =& 1795487 N^8-7087789 N^7+10654130 N^6-5797102 N^5+6828839
   N^4-16594069 N^3\N\\
 &+9651144 N^2+902160 N-1058400~.
\end{align}
Integrals of this type usually contain finite binomial and inverse binomial sums, which 
even may be nested.

In conclusion, we have seen that the methods shown here allow us to obtain analytic 
expressions at general values of $N$ for 3--loop integrals contributing to the massive 
OMEs 
which could not be obtained by other methods before. We continue working on the set of 
integrals that we need in order to obtain all necessary operator matrix elements. 
In particular, we are studying the possibility also to extend the method 
of hyperlogarithms. Application of these methods to the more complicated case
of non-planar integrals are underway. The package {\tt Sigma} and related packages are 
continuously being upgraded to be able to meet the challenges that keep arising in this 
endeavor.

\vspace*{2mm}
\noindent
{\bf Acknowledgment.} ~We would like to thank F.~Brown for discussions. 
The Feynman diagrams have been drawn using {\tt Axodraw}.
This work has 
been supported in part by DFG 
Sonderforschungsbereich Transregio 9, Computergest\"utzte Theoretische 
Teilchenphysik, by the Austrian Science Fund (FWF) grant P203477-N18, by the EU Network 
{\sf LHCPHENOnet} PITN-GA-2010-264564, and ERC Starting Grant PAGAP FP7-257638.


\end{document}